\documentclass[preprin,preprintnumbers,eqsecnum,amsmath,amssymb,nofootinbib]{revtex4}
\usepackage[latin9]{inputenc}
\usepackage{amsmath}

\makeatletter

\usepackage{float}
\usepackage{array}
\usepackage{lipsum}
\usepackage{graphicx}
\usepackage{amsmath,amsthm,amsfonts,amssymb}
\usepackage{graphicx}
\usepackage[english]{babel}
\usepackage{color}
\usepackage{tensor}
\usepackage{esint}
\usepackage[dvips]{epsfig}
\usepackage[dvips]{graphicx}
\usepackage{float}
\usepackage{units}
\usepackage{textcomp}
\usepackage{mathrsfs}
\usepackage{amsmath}
\usepackage[makeroom]{cancel}
\usepackage{amssymb}
\usepackage{amsbsy}
\usepackage{amsfonts}
\usepackage{amssymb,mathrsfs,xcolor}
\usepackage{esint}
\usepackage{array}
\usepackage{graphicx}

\usepackage{hyperref}
\hypersetup{
    colorlinks,
    citecolor=blue,
    filecolor=green,
    linkcolor=purple,
    urlcolor=red,
}

\usepackage{slashed}

\newcommand{\ie}{\begin{equation}}
\newcommand{\fe}{\end{equation}}
\newcommand{\se}{\begin{eqnarray}}
\newcommand{\ff}{\end{eqnarray}}

\usepackage{hyperref}
\hypersetup{colorlinks,breaklinks,
			citecolor=[rgb]{0,0.0,1.0},
            urlcolor=[rgb]{0.0,0.0,1.0},
            linkcolor=[rgb]{0,0.5,0.9}}

\begin{document}

\title{Comment on ``Greybody radiation and quasinormal modes of Kerr-like
black hole in Bumblebee gravity model''}

\author{R. V. Maluf$^{1}$}
\email{r.v.maluf@fisica.ufc.br}

\author{C. R Muniz$^{2}$}
\email{celio.muniz@uece.br}

\affiliation{$^{1}$Universidade Federal do Cear\'{a} (UFC), Departamento de F\'{i}sica, Campus do Pici, Fortaleza - CE, C.P. 6030, 60455-760 - Brazil.\\
 $^{2}$Universidade Estadual do Cear\'{a} (UECE), Faculdade de Educa\c{c}\~{a}o, Ci\^{e}ncias e Letras de Iguatu, Av. D\'{a}rio Rabelo, s/n, Iguatu-CE, 63.500-000 - Brazil.}

\date{\today}

\begin{abstract}
It is shown that the paper ``Greybody radiation and quasinormal modes
of Kerr-like black hole in Bumblebee gravity model'' {[}Eur. Phys.
J. C 81, 501 (2021). arXiv:2102.06303{]} recently published in this
journal is based on an incorrect result obtained by Ding et al. {[}Eur.
Phys. J. C 80, 178 (2020). arXiv:1910.02674{]} for a Kerr-like black
hole solution.
\end{abstract}


\maketitle

\section{Comment}

The paper ``Greybody radiation and quasinormal modes of Kerr-like
black hole in Bumblebee gravity model'' \cite{Kanzi:2021cbg} is an interesting
work that seeks to establish a significant relationship between two
important realms of the current research on astrophysics and high
energy physics, namely, semi-classical gravitation associated to rotating
black holes and Lorentz symmetry breakdown. This latter embodies features
that are beyond the standard model of elementary particles and fundamental
interactions. However, the aforementioned work exhibits inconsistencies
that need to be clarified and corrected, and the present comment seeks
to do this.

Firstly, the complete rotational Kerr-like black hole
solution used in \cite{Kanzi:2021cbg} does not satisfy the equation of motion
for the bumblebee field, as it already was discussed in Ref. \cite{Ding:2020kfr}, although that has not been shown explicitly.
Such a solution exists only in the slow rotation approximation, i.e., when we consider only linear terms in the rotational parameter, $a$. On the other hand, it is possible to show that neither the
modified Einstein equations are satisfied. In fact, after performing a series of manipulations of the gravitational field equations, the authors in \cite{Ding:2019mal} claim to have
found an exact Kerr-like solution for the Einstein-bumblebee theory,
which in Boyer-Lindquist coordinates can be written as
\begin{eqnarray}
ds^{2}=-\Big(1-\frac{2Mr}{\Sigma}\Big)dt^{2}-\frac{4Mra\sqrt{1+\ell}\sin^{2}\theta}{\Sigma}dtd\varphi+\frac{\Sigma}{\Delta}dr^{2}+\Sigma d\theta^{2}+\frac{A\sin^{2}\theta}{\Sigma}d\varphi^{2},\label{bmetric-1-1}
\end{eqnarray}
where $\ell=\varrho b^{2}$ encodes contribution from the Lorentz-violating
effects, and the functions $\Sigma$, $\Delta$ and $A$ are defined as
\begin{equation}
\Sigma(r,\theta)=r^{2}+(1+\ell)a^{2}\cos^{2}\theta,\ \ \Delta(r)=\frac{r^{2}-2Mr}{1+\ell}+a^{2},\;\ A(r,\theta)=\big[r^{2}+(1+\ell)a^{2}\big]^{2}-\Delta(1+\ell)^{2}a^{2}\sin^{2}\theta.
\end{equation}

On the other hand, the equation of motion for the bumblebee field is
\begin{equation}
\nabla^{\mu}b_{\mu\nu}+\frac{\varrho}{\kappa}b^{\mu}R_{\mu\nu}=0,\label{eq:BumblebeeEoM}
\end{equation}
where the bumblebee field strength is $b_{\mu\nu}=\partial_{\mu}b_{\nu}-\partial_{\nu}b_{\mu}$,
with $\varrho$ being a coupling constant, and $\kappa=8\pi G_{N}$. Notice that the original bumblebee field $B_\mu$ was
frozen at its vacuum expectation value (vev) i.e., $B_\mu = b_\mu$ and that its vev has a constant norm
$b_\mu b^\mu= \mp b^2$. These vacuum conditions ensure that the bumblebee field preserves the minimum
of the potential, i.e., $V'=0=V$.

The assumed condition for the bumblebee field is
\begin{equation}
b_{\mu}=(0,b\sqrt{\frac{\Sigma}{\Delta}},0,0),
\end{equation}
such that $b_{\mu}b^{\mu}=b^{2}$ is a positive constant. Thus, the nonzero
components for the bumblebee field strength are $b_{r\theta}=-b_{\theta r}=-b\partial_{\theta}\Sigma/2\sqrt{\Delta\Sigma}$,
and the relevant components obtained from Eq. (\ref{eq:BumblebeeEoM})
are $\nabla^{\theta}b_{\theta r}+\frac{\varrho}{\kappa}b^{r}R_{rr}$
and $\nabla^{r}b_{r\theta}+\frac{\varrho}{\kappa}b^{r}R_{r\theta}$.

For the metric (\ref{bmetric-1-1}), the bumblebee equations of motion
can be explicitly written as
\begin{align}\label{1.5}
 & \nabla^{\theta}b_{\theta r}+\frac{\varrho}{\kappa}b^{r}R_{rr}\nonumber \\
 & =\frac{a^{2}}{2\mathit{b}\kappa\left(r^{2}+a^{2}(1+\ell)\cos^{2}\theta\right)^{3}}\sqrt{\frac{(1+\ell)\left(r^{2}+a^{2}(1+\ell)\cos^{2}\theta\right)}{r^{2}-2 M r+a^{2}(1+\ell)}}\times\nonumber \\
 & \left[a^{2}(1+\ell)\cos^{2}\theta\left(3\ell{}^{2}-4 b^{2}\kappa(1+\ell)+\ell^{2}\cos2\theta\right)+r^{2}\left(\ell^{2}-b^{2}\kappa(1+\ell)\right)\left(1+3\cos2\theta\right)\right]\nonumber \\
 & \neq0,
\end{align}
and
\begin{align}\label{1.6}
 & \nabla^{r}b_{r\theta}+\frac{\varrho}{\kappa}b^{r}R_{r\theta}\nonumber \\
 & =\frac{a^{2}b\sin2\theta}{4\left(r^{2}+a^{2}(1+\ell)\cos^{2}\theta\right)^{3}}\sqrt{\frac{(1+\ell)\left(r^{2}+a^{2}(1+\ell)\cos^{2}\theta\right)}{r^{2}-2 M r+a^{2}(1+\ell)}}\times\nonumber \\
 & \left[r\left(a^{2}(1+\ell)(-5+\cos2\theta)+10 Mr-4r^{2}\right)-2a^{2}(1+\ell) M \cos^{2}\theta\right]\nonumber \\
 & \neq0,
\end{align} where we replace $\varrho=\ell/b^{2}$. In this way, the solution
presented in \cite{Ding:2019mal} fails to solve exactly the equation
of motion for the bumblebee field. However, in the slowly rotating
limit, i.e., $a^{2}\rightarrow0$, these equations are fulfilled.
They become
\begin{align}
 & \nabla^{\theta}b_{\theta r}+\frac{\varrho}{\kappa}b^{r}R_{rr}\nonumber \\
 & =\frac{a^{2}(\ell^{2}-b^{2}(1+\ell)\kappa)(1+3\cos2\theta)}{2 b\kappa r^{4}}.\sqrt{\frac{(1+\ell)r}{r-2 M}}+\mathcal{O}(a^{3})\nonumber \\
 & \cong0\ \ \ \ \mbox{when}\ \ \  a^{2}\rightarrow 0,
\end{align}
and
\begin{align}
 & \nabla^{r}b_{r\theta}+\frac{\varrho}{\kappa}b^{r}R_{r\theta}\nonumber \\
 & =-\frac{a^{2}b(2r-5 M)\sin2\theta}{2r^{4}}.\sqrt{\frac{(1+\ell)r}{r-2 M}}+\mathcal{O}(a^{3})\nonumber \\
 & \cong0\ \ \ \ \mbox{when} \ \ \ a^{2}\rightarrow 0,\\
\nonumber
\end{align} where $\mathcal{O}(\mathit{a}^{3})$ denotes terms of order greater
than or equal to $a^{3}$. It is worth mentioning that for a small $b$ limit, the equations (\ref{1.5}) and (\ref{1.6}) should at least be proportional to $b$. Indeed, these quantities take the form
\begin{align}
 & \nabla^{\theta}b_{\theta r}+\frac{\varrho}{\kappa}b^{r}R_{rr}\nonumber \\
 & = -\frac{ba^{2}\left[4a^{2}\cos^{2}\theta+r^{2}\left(1+3\cos2\theta\right)\right]}{2\left(r^{2}+a^{2}\cos^{2}\theta\right)^{3}}\sqrt{\frac{r^{2}+a^{2}\cos^{2}\theta}{r^{2}-2Mr+a^{2}}}+\mathcal{O}(b^{2}),
 \end{align}
and
\begin{align}
 & \nabla^{r}b_{r\theta}+\frac{\varrho}{\kappa}b^{r}R_{r\theta}\nonumber \\
 & =-\frac{ba^{2}\left[4r^{3}-10Mr^{2}+2a^{2}M\cos^{2}\theta+a^{2}r(5-\cos2\theta)\right]\sin2\theta}{4\left(r^{2}+a^{2}\cos^{2}\theta\right)^{3}}\sqrt{\frac{r^{2}+a^{2}\cos^{2}\theta}{r^{2}-2Mr+a^{2}}}+\mathcal{O}(b^{2}),
\end{align}and essentially, the equation of motion (\ref{eq:BumblebeeEoM}) is not satisfied at the small $b$ limit either.
Such results are in agreement with those
presented in \cite{Ding:2020kfr}, and similarly, it can be shown explicitly
(using the computer algebra program \cite{Martin-Garcia:2007bqa} ) that the Einstein-bumblebee
gravitational equations are also not satisfied by the solution (\ref{bmetric-1-1}).
Thus, the existence of a full rotating black hole solution for the
Einstein-bumblebee theory remains an open issue.

From the foregoing, the authors of reference \cite{Kanzi:2021cbg} spoiled
their work from the beginning, and we fear it is irremediable. However,
the physical implications of the studied model might be considered
if the original results are fitted to the slow rotating solution approximation.

\begin{acknowledgments}
The authors thank the Funda\c{c}\~{a}o Cearense de Apoio ao Desenvolvimento
Cient\'{i}fico e Tecnol\'{o}gico (FUNCAP), the Coordena\c{c}\~{a}o de Aperfei\c{c}oamento de Pessoal de N\'{i}vel Superior (CAPES), and the Conselho Nacional de Desenvolvimento Cient\'{i}fico e Tecnol\'{o}gico (CNPq), Grants no 307556/2018-2 (RVM) and no 308979/2018-4 (CRM) for financial support.

\end{acknowledgments}


\appendix*
\section{Some Quantities}

In this appendix, we list out the nonzero components of the Ricci tensor and the covariant derivative of $b_{\mu\nu}$ to make checking equations (\ref{1.5}) and (\ref{1.6}) easier for the readers. For the metric (\ref{bmetric-1-1}), they take the explicit form:
\begin{equation}
R_{rr}=\frac{\ell a^{2}(1+\ell)\left[a^{2}(1+\ell)\cos^{2}\theta(3+\cos2\theta)+r^{2}(1+3\cos2\theta)\right]}{2\left(r^{2}-2Mr+a^{2}(1+\ell)\right)\left(r^{2}+a^{2}(1+\ell)\cos^{2}\theta\right)^{2}},
\end{equation} which at the limit $\ell\rightarrow 0$ it vanishes, and additionally, we verify that the component $R_{r\theta}$ is identically null. These results agree that the Kerr metric is Ricci flat in the absence of the Lorentz violation since it is the solution of the vacuum Einstein equations. Besides, the relevant covariant derivatives are given by
\begin{align}
&\nabla^{\theta}b_{\theta r}	=-\frac{a^{2}b(1+\ell)}{2\left(r^{2}+a^{2}(1+\ell)\cos^{2}\theta\right)^{3}}\sqrt{\frac{(1+\ell)\left(r^{2}+a^{2}(1+\ell)\cos^{2}\theta\right)}{r^{2}-2Mr+a^{2}(1+\ell)}}\times\nonumber\\
 &\left[4a^{2}(1+\ell)\cos^{2}\theta+r^{2}(1+3\cos2\theta)\right],
 \end{align}
and
\begin{align}
  &\nabla^{r}b_{r\theta} =\frac{a^{2}b\sin2\theta}{4\left(r^{2}+a^{2}(1+\ell)\cos^{2}\theta\right)^{3}}\sqrt{\frac{(1+\ell)\left(r^{2}+a^{2}(1+\ell)\cos^{2}\theta\right)}{r^{2}-2 M r+a^{2}(1+\ell)}}\times\nonumber \\
 & \left[r\left(a^{2}(1+\ell)(-5+\cos2\theta)+10 Mr-4r^{2}\right)-2a^{2}(1+\ell) M \cos^{2}\theta\right].
 \end{align}


\end{document}